\documentclass{aa} 
\usepackage{etex}
\usepackage{color}
\usepackage{graphicx}
\usepackage{txfonts}
\newcommand{\AIPS}{$\mathcal{A}\mathcal{I}\mathcal{P}\mathcal{S}$}
\begin{document} 

   \title{Discovery of off-axis jet structure of TeV Blazar Mrk 501\\with mm-VLBI}
   \author{S. Koyama\inst{1},
        M. Kino\inst{2},
        M. Giroletti\inst{3},
        A. Doi\inst{4},
        G. Giovannini\inst{3,5},
        M. Orienti\inst{3},
        K. Hada\inst{3,6},\\
        E. Ros\inst{1,7,8},
        K. Niinuma\inst{9},
        H. Nagai\inst{6},
        T. Savolainen\inst{1,10},
         T. P. Krichbaum\inst{1}, 
         and
         M. \'{A}. P\'{e}rez-Torres\inst{11,12}}
   \institute{
        \inst{1}~Max-Planck-Institut f\"{u}r Radioastronomie, Auf dem H\"{u}gel 69, D-53121 Bonn, Germany\\
        \email{skoyama@mpifr-bonn.mpg.de}\\
        \inst{2}~Korea Astronomy and Space Science Institute, 776 Daedeokdae-ro, Yuseong-gu, Daejeon, 305-348, Republic of Korea\\
        \inst{3}~INAF Istituto di Radioastronomia, via Gobetti 101, I-40129 Bologna, Italy\\
        \inst{4}~Institute of Space and Astronautical Science, Japan Aerospace Exploration Agency, 3-1-1 Yoshinodai, Chuou-ku, Sagamihara, Kanagawa 252-5210, Japan\\
        \inst{5}~Dipartimento di Astronomia, Universit\`{a} di Bologna, via Ranzani 1, I-40127 Bologna, Italy\\
        \inst{6}~National Astronomical Observatory of Japan, 2-21-1 Osawa, Mitaka, Tokyo 181-8588, Japan\\
        \inst{7}~Observatori Astron\`{o}mic, Universitat de Val\`{e}ncia, E-46980 Paterna, Val\`{e}ncia, Spain\\
        \inst{8}~Departament d'Astronomia i Astrof\'{i}sica, Universitat de Val\`{e}ncia, E-46100 Burjassot, Val\`{e}ncia, Spain\\
        \inst{9}~Graduate School of Science and Engineering, Yamaguchi University, Yamaguchi 753-8511, Japan\\
        \inst{10}~Aalto University Mets\"{a}hovi Radio Observatory, FIN-02540 Kylm\"{a}l\"{a}, Finland\\
        \inst{11}~Instituto de Astrof\'{i}sica de Andaluc\'{i}a, Glorieta de la Astronom\'{i}a, s/n, E-18008 Granada, Spain\\
        \inst{12}~Centro de Estudios de la F\'{i}sica del Cosmos de Arag\'{o}n, E-44001 Teruel, Spain}
   \date{Received, 17 May, 2015; accepted; 7 December, 2015}
 
  \abstract
   {High-resolution millimeter wave very-long-baseline interferometry (mm-VLBI) is an ideal tool for probing the structure at the base of extragalactic jets in detail.
   The TeV blazar Mrk~501 is one of the best targets among BL Lac objects for studying the nature of off-axis jet structures because it shows different jet position angles at different scales.}
   {The aim of this study is to investigate the properties of the off-axis jet structure 
   through high-resolution mm-VLBI images at the jet base and
   physical parameters such as kinematics, flux densities, and spectral indices.}
   {We performed Very Long Baseline Array (VLBA) observations 
   over six epochs from 2012 February to 2013 February at 43\,GHz.
   Quasi-simultaneous Global Millimeter VLBI Array (GMVA) observations at 86\,GHz
   were performed in May 2012.}
   {We discover a new jet component at the northeast direction from the core 
   in all the images at 43 and 86\,GHz. 
   The new component shows the off-axis location from the persistent jet extending to the southeast.
   The 43\,GHz images reveal that the scattering of the positions of the NE component is within $\sim0.2$\,mas.
   The 86\,GHz data reveals a jet component located 0.75\,mas southeast of the radio core.
   We also discuss the spectral indices 
   between 43 and 86\,GHz, where 
   the northeast component has steeper spectral index 
   and the southeast component has comparable or flatter index
   than the radio core does.}
   {}

   \authorrunning{S. Koyama et al.}
   \maketitle

\section{Introduction}
Relativistic outflows (jets) in active galactic nuclei (AGNs)
are ultimately powered by gravitational energy
of supermassive black holes lying in the center of galaxies
\citep[e.g.,][]{Begelman:1984aa}.
Probing structures and kinematics of the inner parts of jets
is a fundamental issue for understanding jet physics.
The intensive systematic Very Long Baseline Array (VLBA)
monitoring of relativistic outflows
in a sample of over 100 AGN jets at 15\,GHz has been conducted
by the Monitoring Of Jets in Active galactic nuclei with VLBA Experiments (MOJAVE) project
\citep[e.g.,][]{Kellermann:1998,Lister:2009lr,Lister:2013,Homan:2009,Homan:2015}.
The MOJAVE data have revealed that 
a large fraction of jets at milliarcsecond scales show evidence of curved motion.

Recently, a few very-long-baseline interferometry (VLBI) observations at frequencies higher than 43\,GHz
have revealed the inner structure of the curved jets 
and their kinematics at submilliarcsecond scales.
Monthly monitoring of blazars with the VLBA at 43\,GHz
reveals large swings of the jet position angle ($PA$) in their innermost regions
(e.g., OJ~287: \citealt{Agudo:2012}, as a part of the Boston University Blazar Project).
\cite{Lu:2013} find an unusual jet $PA$ in the quasar 3C~279
by combining Event Horizon Telescope observations at 230\,GHz
and VLBA observations at 43\,GHz at $\sim$0.2\,mas ($\sim$1\,pc) scales.
The new jet $PA$ is almost 
perpendicular to the persistent jet axis
at a scale of a few pc observed
at or lower than 43\,GHz
(hereafter called ``off-axis jet structure'').
Thus,
high-resolution
VLBI observations at high frequencies ($\ge$43\,GHz)
are important
to explore such off-axis jet structures.
Off-axis jet structures at the base of radio jets
could be related to the properties of the jet formation regions;
however,
their nature is still poorly understood.

The TeV blazar Mrk~501 is one of the best BL Lac objects
for studying the innermost jet thanks to its proximity ($z\,=\,0.034$) and brightness.
The jet $PA$s of this source 
show different apparent directions on different angular scales,
changing smoothly in counter-clockwise direction.
Low-frequency observations have revealed that
the jet $PA$ (defined positive north through east)
is $\sim45^{\circ}$ at $r>20$\,mas from the core,
and $\sim100^{\circ}$ at $2<r<20$\,mas
\citep{Giroletti:2004,Giroletti:2008}.
VLBA images at 43\,GHz 
have shown a clear limb-brightened structure toward 
$PA\sim150^{\circ}$ (labeled eastern limb) and $PA\sim175^{\circ}$ (labeled western limb)
at $0.5<r<2$\,mas
\citep{Piner:2009, Piner:2010}. 
The Global Millimeter VLBI Array (GMVA) observations at 86\,GHz
show a jet feature toward $PA\sim172^{\circ}$ at 0.73\,mas
and the inner 0.1\,mas jet limbs at $PA\sim-135^{\circ}$ and $PA\sim144^{\circ}$
\citep{Giroletti:2008}.

In this paper, we report on the sub-pc to pc scale structure of Mrk~501 obtained 
with VLBA 43\,GHz and GMVA 86\,GHz observations in 2012--2013,
focusing on a newly found off-axis jet structure.
In Sect.~2 we describe the observations; 
in Sect.~3 we present the results; and in Sect.~4 we give a discussion of our findings.
Throughout this paper, we adopt the following 
cosmological parameters:
$H_{0}\,=\,71~ {\rm km~s^{-1} Mpc^{-1}}$,
$\Omega_{\rm M}\,=\,0.27$, and    
$\Omega_{\rm \Lambda}\,=\,0.73$ \citep{Komatsu:2009}, or
$1\,{\rm mas}\,=\,0.66$~pc for Mrk~501.


\section{Observations and data reductions}
\subsection{VLBA}

In Table~\ref{vlba43},
we summarize the details of our VLBI observations.
Six epochs of observations were carried out at 43\,GHz between 2012 February and 2013 February with ten VLBA stations:
Kitt Peak (KP), Saint Croix (SC), Fort Davis (FD), Owens Valley (OV), Mauna Kea (MK), Hancock (HN), Pie Town (PT),
North Liberty (NL), Brewster (BR), and Los Alamos (LA).
The central frequency was 43.212\,GHz, with eight sub-bands (IFs) of 16\,MHz bandwidth each.
Left hand circular polarization (LHCP) was recorded at a bit-rate of 512~Mbps by Mark 5 disk systems,
and was correlated at the National Radio Astronomy Observatory (NRAO) VLBA-DiFX software correlator \citep{Deller:2011}.
In some epochs, one or more antennas did not work properly because of technical problems or weather condition (see Table~\ref{vlba43}).
For each epoch, the total on-source time for Mrk~501 was approximately one hour,
distributed over a four-hour long track.

Initial data calibration was performed using the Astronomical Image Processing System (\AIPS) 
software package developed by the NRAO.
A priori amplitude calibration was done based on the measurements of the system noise temperature 
during the observations and the elevation-dependent antenna gain provided by each station.
In this process, we also applied opacity correction due to the atmospheric attenuation, 
assuming that the time variation of the opacity is not significant during each observation.
Phase and delay offset between different sub-bands were solved by using the calibrator 3C~345.
Fringe fitting was performed with the \AIPS 
~task {\sc fring} on Mrk~501 by averaging over all the IFs.
Imaging and self-calibration were performed using the Difmap software package \citep{Shepherd:1997}.
The final images were produced after iterations of CLEAN, phase, and amplitude self-calibration processes.

\begin{figure*}
\centering
\includegraphics[width=11cm]{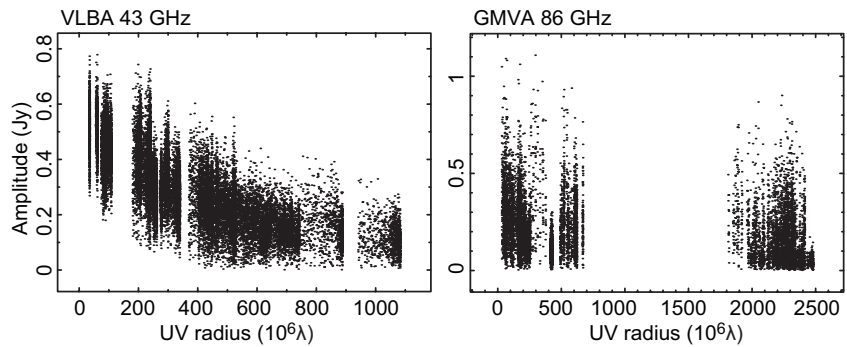}
\caption{Visibility amplitude vs. $(u, v)$-radius
for the third epoch VLBA observations at 43\,GHz ({\it left panel})
and for the GMVA observations at 86\,GHz ({\it right panel}).}
\label{rad}
\end{figure*}

The ($u,v$) range is 32--1082~M$\lambda$ for the third epoch as an example (Fig.~\ref{rad} {\it left}),
and its maximum is 1242~M$\lambda$ for the second epoch.

\begin{table*}
\caption{Journal of VLBI observations}
\label{vlba43}
\begin{center}
\begin{tabular}{cccccccccc}
\hline\hline
Epoch&Observing&Frequency &Map peak&\multicolumn{3}{c}{Beam size\tablefootmark{$\mathrm a$}}&1$\sigma$&Notes\\ 
&Code&(GHz)&$\mathrm{(mJy~beam^{-1})}$&(mas)&(mas)&(deg)&$\mathrm{(mJy~beam^{-1})}$&\\
\hline
2012 Feb 12&BK172A &43 & 209& 0.46 &0.12& $-$22.4& 1.25 &       no KP, SC, FD\\2012 Mar 16 & BK172B &43 & 165 & 0.33 & 0.12 & $-$15.0 & 0.98 & \\
2012 May 06 & BK172C &43 & 215 & 0.40 & 0.12 & $-$17.2 & 1.30 & OV, SC\tablefootmark{$\mathrm b$}\\ 
2012 Jun 11 & BK172D &43 & 267 & 0.35 & 0.17 & ~~12.8 & 0.96 & no MK\\
2013 Jan 18 & BK172E &43 & 148 & 0.35 & 0.13 & $-$19.6 & 0.85 & HN\tablefootmark{$\mathrm b$}\\
2013 Feb 15 & BK172F &43 & 144 & 0.33 & 0.12 & $-$22.0 & 1.06 & FD, PT\tablefootmark{$\mathrm b$}\\
\hline
2012 May 19 & GG072 &86 & ~~99 & 0.21 & 0.04 & $-$5.11 & 1.93  & see details in \S\ref{gm}\\
\hline
\end{tabular}
\end{center}
~\tablefoottext{a}{Major axis, minor axis, and position angle of synthesized beam.}
\tablefoottext{b}{The stations were affected by significant weather or technical problems.}
\end{table*}

\begin{figure*}
\centering
\includegraphics[width=13cm]{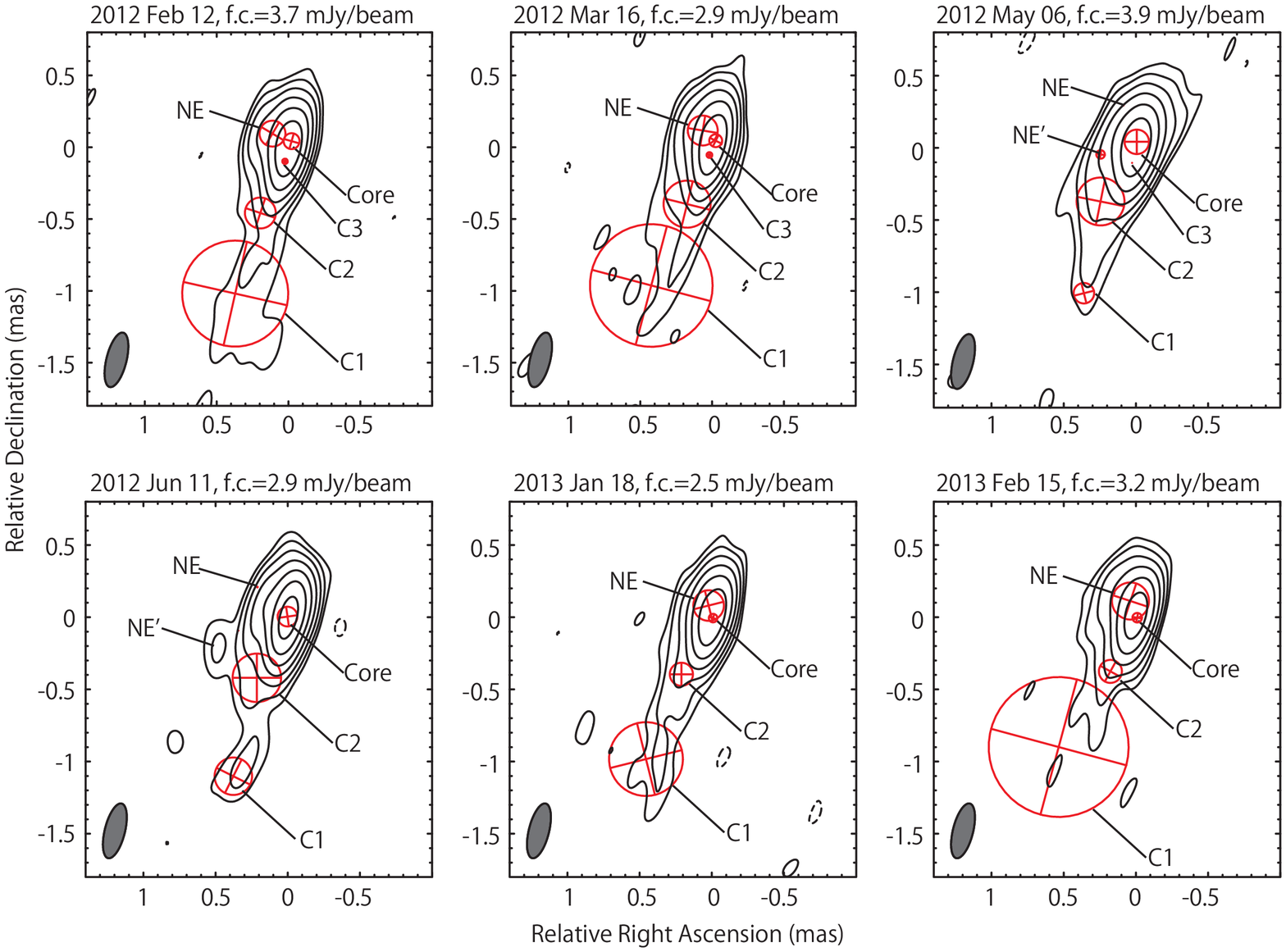}
\caption{Uniform weighted VLBA 43\,GHz CLEAN images with fitted circular Gaussian components.
The restoring beam is 0.39\,mas $\times$ 0.14\,mas in $PA=-14^{\circ}$, plotted in the bottom left corner.
Observing date and the first contour (f.c.) are plotted above each map.
The first contours are all set to three times the rms noise level of each map, increasing by a factor of 2. }
\label{res01835}
\end{figure*}

\subsection{GMVA}\label{gm}

We observed Mrk~501 on 2012 May 19 with the GMVA at 86\,GHz,
which was 13~days after the third epoch of VLBA 43\,GHz observations.
The central frequency was 86.198\,GHz, with eight~IFs of 16\,MHz bandwidth each. The
LHCP was recorded at a bit-rate of 512~Mbps by Mark 5 disk systems,
and was correlated at the Bonn DiFX software correlator.
The participating instruments were four European telescopes 
(Effelsberg (EF), the Plateau de Bure interferometer (PB), Onsala (ON), and Yebes (YB)), 
and seven VLBA stations (all except SC, HN, and MK).
We observed the calibrator 3C~345 and the target with a time cycle of 16 minutes,
where two minutes were allocated for calibrator 3C~345, five minutes for target Mrk~501,
and the remaining time for antenna pointing and calibration.
The total observation time was approximately 12\,hrs (UT 0/20:00--1/08:30).
The European telescopes observed for the first $\sim$9\,hrs, 
and the VLBA joined in for the last $\sim$6\,hrs.
The telescopes at EF and OV failed
owing to system problems 
(low signal-to-noise ratio (S/N), and power supply problems, respectively).

The calibrator 3C~345 was detected with a S/N 
$>7$ on all baselines. 
From the fringe fitting of 3C~345, 
we determined single-band delay offsets using good scans 
and applied them to the whole data set.
We performed global fringe fitting on 3C~345 over all the IFs to determine the phase, delay, and rate, 
using a solution interval of two minutes.
CLEAN and phase self-calibration were performed iteratively, 
and one amplitude self-calibration for 3C~345 was applied using 
the {\sc gscale} command of Difmap.
We obtained an image of 3C~345 
and found the structure within 2\,mas from the core
to be in agreement with the archival 43\,GHz VLBA image
taken around a week after our GMVA session 
(on 2012 May 27)\footnote[1]{http://www.bu.edu/blazars/VLBAproject.html}.%
The total GMVA 86\,GHz cleaned flux of 3C~345 was 1.7\,Jy,
which is approximately 50\% of the single-dish flux at 86\,GHz
\footnote[2]{http://www3.mpifr-bonn.mpg.de/div/vlbi/fgamma/results.html}. %
Since there are only small differences ($\sim$5\%) between the single-dish fluxes and the total VLBA fluxes
at 15\,GHz\footnote[3]{http://www.physics.purdue.edu/astro/MOJAVE/allsources.html} 
and 43\,GHz$^{1}$, 
we attribute this much larger difference to a significant flux loss for the GMVA 86\,GHz.
Such flux losses can sometimes be seen in the images obtained with GMVA at 86\,GHz
\citep[e.g.,][]{Rani:2015a}.
The possible reasons could be antenna pointing error,
unmodeled antenna efficiency, and missing flux from extended structure.
To derive a scaling factor of 3C~345 for the GMVA 86\,GHz,
we estimated the extended flux at 15\,GHz and 43\,GHz
as the difference between the single-dish fluxes and the total cleaned VLBI fluxes, 
and extrapolated the value of the extended flux at 86 GHz 
by using the spectral index ($\alpha\sim-0.6$ as determined between 15\,GHz and 43\,GHz). 
We then estimated the expected GMVA 86\,GHz flux as $\sim3.5$\,Jy,
and determined the scaling factor to be around 2.1 for 3C~345.
Since the source compactness varies between sources
\citep[e.g.,][]{Lee:2008}, 
we give the estimate of the scaling factor for Mrk~501 at the end of this section.

At this stage, it was then possible to fringe fit Mrk~501 itself 
using the fringe solutions of 3C~345, averaging over all the IFs. 
For fringe fitting, we used a solution interval of five minutes
because the S/N of the solutions increased as we increased the solution intervals 
by 0.5 minutes (0.5--5 minutes).
Since 3C~345 is close to Mrk~501 (separation of $2.09^{\circ}$), 
most of the phase variations were corrected by 3C~345
and the residual delay and phase rate for Mrk~501 became quite small. 
Therefore, we set a small search window of a few tens of nsec and mHz in delay and rate, respectively, 
with a low S/N threshold. 
The overall ($u,v$) range is 29--2476\,M$\lambda$ 
with sparse coverage in the interval 668--1809\,M$\lambda$ (Fig.~\ref{rad} {\it right}).
After some iterations of fringe fitting,
we find smooth delay and phase rate solutions clustering each other in the following conditions.
Fringes were found between European stations (setting S/N cutoff~2.8, maximum S/N$\sim$7.0) 
and between VLBA telescopes (setting S/N cutoff~2.5, maximum S/N$\sim$6.0).
Fringes of transatlantic baselines were only detected between PB and KP, and PB and LA,
during several good scans
(setting S/N cutoff~2.5, maximum S/N$\sim$3.0).
Although the solutions on the long baselines might be less significant
(they may lead to the overestimation of the correlated flux and compactness), 
the emission on the short baselines is clearly detected.
In any case, we edited out the solutions that were obviously bad with the \AIPS 
~task {\sc snedt} and we 
subsequently frequency averaged the data.
Gaussian model fitting and phase self-calibration were performed in Difmap.
Owing to the low S/N of the visibilities,
we did not perform amplitude self-calibration.

We then derived the scaling factor for the GMVA image of Mrk~501.
The total modeled flux of Mrk~501 by the GMVA at 86\,GHz was 286\,mJy,
which is approximately 40\% of the single-dish flux ($\sim750$\,mJy) at 86\,GHz$^{2}$,
obtained around a week before the GMVA observations.
On the other hand, at 15\,GHz$^{3}$ and 43\,GHz$^{1}$,
the total cleaned VLBI fluxes are around 60\% of the single-dish measurement,
such that there might be some flux loss for the GMVA flux.
We derived the spectral index of the extended flux to be $\alpha\sim-0.4$ in the same manner as 3C~345,
then estimated the expected GMVA 86\,GHz flux as $\sim510$\,mJy,
and finally determined the scaling factor to be around 1.8 for Mrk~501.
%

\section{Results}
\subsection{VLBA results}\label{vlbares}
\subsubsection{Images}~\label{IM}

Figure~\ref{res01835} shows the total-intensity CLEAN images 
obtained with the VLBA at 43\,GHz as contours.
The rms noise level on our images is approximately 1\,$\rm mJy~beam^{-1}$, 
a factor of a few times higher than the noise level expected theoretically 
($0.32\sim0.47\,{\rm mJy~beam^{-1}}$ estimated from VLBA observational status summary).
The rms level is comparable to or slightly higher than the noise level 
on the images of Mrk~501 in \cite{Piner:2009},
whose observing frequency, aperture, settings, and on-source time (40 minutes) 
are similar to ours.

All the images show a bright core
and a one-sided jet structure elongated toward the southeast.
The core and one-sided limb-brightened southeast jet
have been observed in previous images, and we give a 
detailed comparison in \S\ref{AP}.
Our most remarkable finding is the emergence of 
new emission toward the northeast and to the east of the core in 
all six images.
Compared to the dense contours west of the core, 
the contours east of the core are clearly sparse and indicate the presence of emission from an additional component.
We searched the website of Boston University's blazar monitoring program$^{1}$ for images of similar quality.
All the images available between 2012 October 19 and 2014 December 5
are in good agreement with our images,
showing emission elongated toward the northeast to the east of the core.
In the VLBA 43\,GHz image obtained on 2011 September 24,
the northeast emission was not seen,
possibly due to the lack of dynamic range.

\cite{Piner:2009,Piner:2010} presented VLBA 43\,GHz images 
between 2005 and 2009 of a quality comparable to ours;
however,
clear emission elongated northeast of the core was not detected.
Therefore,
the northeast emission has emerged at least since the first epoch of our observations (2012 February 12),
and the emergence might be related to a $\gamma$-ray flare in 2011 October \citep{Bartoli:2012}.

\subsubsection{Model fitting}\label{modsec}

In order to parametrize the brightness distribution of the jet, 
we used the task {\sc modelfit} in Difmap.
In the model-fit process, 
we used a set of circular Gaussian model components 
to fit the visibility data in the ($u,v$) plane.
In Fig.~\ref{res01835}, the model components are overlaid 
on the CLEAN images.
Table~\ref{gaus} lists the resulting model-fit parameters 
at every observing epoch.

Overall structures are mostly represented by four to six model components.
We note that without
a model component northeast of the core,
there were brighter residuals at the northeast of the core than at the southeast of the core.
For the one month separation data,
i.e. between 2012 February and March, between 2012 May and June, 
and between 2013 January and February,
we repeated the model fitting of the latter epochs 
using the model components from the former epochs.
For the data with longer than two months separation,
we performed the model fitting procedure independently
due to the possible structural changes.
In the component identification process, 
we identified the brightest, most compact, and innermost feature as the radio core,
and the northeast component as NE.
Hereafter, the other components are as follows: 
C1 as the outermost component, 
C2 as the second outermost component,
and C3 as the third outermost jet component 
toward the southeast.
The jet extends to roughly 1.1\,mas southeast of the core. 
In the third and fourth epoch images, 
we find the component labeled NE$'$,
which elongates southeast 
of the NE component.

The uncertainties of the model-fit parameters presented in Table~\ref{gaus}
are estimated by following the methods described in \cite{Lico:2012} and \cite{Blasi:2013}.
The position errors of each component 
are estimated using the ratio of the size of each component to the S/N.
However, most of the components 
are smaller than the beam size of the image,
such that the position errors are replaced by a conservative value equal to 10\% of the beam size \citep{Lister:2009lr, Orienti2011}.
The flux density uncertainties are estimated considering the addition in quadrature of
a calibration error of about 10\% and a statistical error given by three times the rms noise of the map.

\begin{table*}
\begin{center}
\caption{Gaussian models for the calibrated visibilities of VLBA 43\,GHz and GMVA 86\,GHz data}
\label{gaus}
\begin{tabular}{ccccccccccccc}
\hline\hline
Epoch& Component & $S\pm\sigma_{S}$\tablefootmark{a} & $r\pm\sigma_{r}$\tablefootmark{b} & 
$PA\pm\sigma_{PA}$\tablefootmark{b} & $a\pm\sigma_{a}$\tablefootmark{c} \\
&&(mJy) & (mas) & (deg)  & (mas) \\
\hline
2012 Feb 12\tablefootmark{d} & Core & 188 $\pm$ 19.4 & ... &  ... & 0.11 $\pm$ 0.05\\
& NE & 73 $\pm$ 8.6 & 0.14 $\pm$ 0.01 & 68.5 $\pm$ 4.6 & 0.18 $\pm$ 0.02\\
& C3 & 83 $\pm$ 9.5 & 0.15 $\pm$ 0.03 & 161.9 $\pm$ 16.7 & 0.04 $\pm$ 0.09\\
& C2 & 10 $\pm$ 4.7 & 0.54 $\pm$ 0.05 & 156.5 $\pm$ 4.8 & 0.21 $\pm$ 0.05\\
& C1 & 54 $\pm$ 7.0 & 1.13 $\pm$ 0.11 & 159.7 $\pm$ 3.9 & 0.73 $\pm$ 0.22\\
2012 Mar 16\tablefootmark{d}  & Core & 106 $\pm$ 11.1 & ...  & ... & 0.09 $\pm$ 0.06\\
& NE & ~~98 $\pm$ 11.1 & 0.11 $\pm$ 0.01 & 51.8 $\pm$ 6.4 & 0.21 $\pm$ 0.03\\
& C3 & 77 $\pm$ 8.4 & 0.11 $\pm$ 0.03 & 156.4 $\pm$ 15.8 & 0.04 $\pm$ 0.06\\
& C2 & 24 $\pm$ 4.1 & 0.48 $\pm$ 0.03 & 155.8 $\pm$ 3.6 & 0.33 $\pm$ 0.06\\
& C1 & 43 $\pm$ 5.4 & 1.10 $\pm$ 0.15 & 156.0 $\pm$ 7.4 & 0.85 $\pm$ 0.30\\
2012 May 06\tablefootmark{d}  & Core & 335 $\pm$ 33.8 & ... &  ... & 0.17 $\pm$ 0.06\\
& NE& 15 $\pm$ 5.1 & 0.27 $\pm$ 0.02 & 20.1 $\pm$ 3.9 & ... \\ 
& NE$'$ & 25 $\pm$ 5.4 & 0.27 $\pm$ 0.01 & 109.1 $\pm$ 3.1 & 0.06 $\pm$ 0.03\\
& C3 & 50 $\pm$ 6.9 & 0.15 $\pm$ 0.04 & 167.5 $\pm$ 14.5 & ... \\ 
& C1 & 17 $\pm$ 5.1 & 1.11 $\pm$ 0.04 & 160.7 $\pm$ 2.1 & 0.14 $\pm$ 0.08\\
2012 Jun 11\tablefootmark{d}  & Core & 353 $\pm$ 35.5 & ... & ... & 0.14 $\pm$ 0.07 \\
& NE & 15 $\pm$ 3.9 & 0.29 $\pm$ 0.03 & 45.6 $\pm$ 5.0 & ... \\
& NE$'$ & 11 $\pm$ 3.8 & 0.35 $\pm$ 0.02 & 105.9 $\pm$ 2.8 & ... \\
& C2 & 33 $\pm$ 4.9 & 0.45 $\pm$ 0.03 & 153.4 $\pm$ 2.8 & 0.34 $\pm$ 0.05\\
& C1 & 18 $\pm$ 4.0 & 1.17 $\pm$  0.05 & 161.3 $\pm$ 1.3 & 0.26 $\pm$ 0.14\\
2013 Jan 18\tablefootmark{d}  & Core & 130 $\pm$ 13.4 & ...  & ... & 0.06 $\pm$ 0.05\\
& NE & 88 $\pm$ 9.4 & 0.09 $\pm$ 0.02 & 19.6 $\pm$ 11.6 & 0.21 $\pm$ 0.04\\
& C2 & 17 $\pm$ 3.7 & 0.45 $\pm$ 0.03 & 150.6 $\pm$ 4.2 & 0.16 $\pm$ 0.07\\
& C1 & 21 $\pm$ 3.9 & 1.08 $\pm$ 0.10 & 154.5 $\pm$ 0.10 & 0.51 $\pm$ 0.20\\
2013 Feb 15\tablefootmark{d}  & Core & 148 $\pm$ 15.2 & ... & ... & 0.07 $\pm$ 0.05 \\
& NE & 109 $\pm$ 11.6 & 0.12 $\pm$ 0.02 & 22.0 $\pm$ 7.4 & 0.26 $\pm$ 0.03\\
& C2 & 17 $\pm$ 4.2 & 0.42 $\pm$ 0.03 & 153.5 $\pm$ 4.4 & 0.16 $\pm$ 0.06\\
& C1 & 53 $\pm$ 6.6 & 1.05 $\pm$ 0.25 & 148.7 $\pm$ 10.8 & 0.97 $\pm$ 0.04\\
\hline
2012 May 19\tablefootmark{e} & Core & 235 $\pm$ 71.4 & ... & ... & 0.06 $\pm$ 0.05\\ 
& NE & 228 $\pm$ 69.2 & 0.11 $\pm$ 0.02 & ~~49.3 $\pm$ 19.2 & 0.67 $\pm$ 0.04\\
& SE & ~~23 $\pm$ 12.8 & 0.75 $\pm$ 0.02 & 155.8 $\pm$ 3.3 & 0.07 $\pm$ 0.04\\
& SE$'$ & ~~45 $\pm$ 17.2 & 1.28 $\pm$ 0.02 & 159.5 $\pm$ 2.0 & 0.17 $\pm$ 0.05\\
\hline
\end{tabular}
\end{center}
~\tablefoottext{a}{Flux density and estimated errors.}
\tablefoottext{b}{$r$ and $PA$s  are the polar coordinates of the component's center with respect to the core. The $PA$s are measured from North through East. $\sigma_{r}$ are estimated errors in the component position. $\sigma_{PA}$ are estimated errors in the $PA$ of the component.}
\tablefoottext{c}{The FWHM of the radius of the circular Gaussian component and estimated errors in the component size.}
\tablefoottext{d}{The VLBA 43\,GHz data.}
\tablefoottext{e}{The GMVA 86\,GHz data.}
\end{table*}

\subsubsection{Angular separation and position angle of individual components}\label{AP}

\begin{figure}
\centering
\includegraphics[width=7cm]{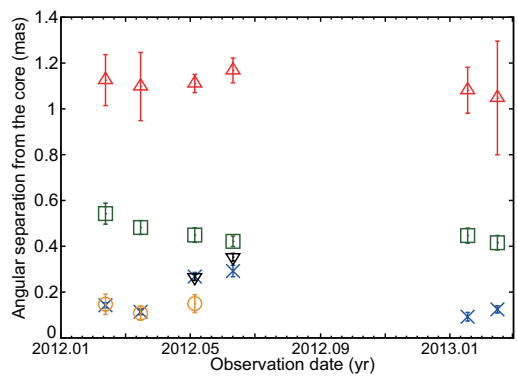}
\caption{Angular separation from the core to each component:  C1 (blank red triangles), C2 (blank green squares), C3 (blank orange circles), NE (blue crosses), and NE$'$ (blank black inverted triangles).}
\label{vlbarad}
\end{figure}

\begin{figure}
\centering
\includegraphics[width=7cm]{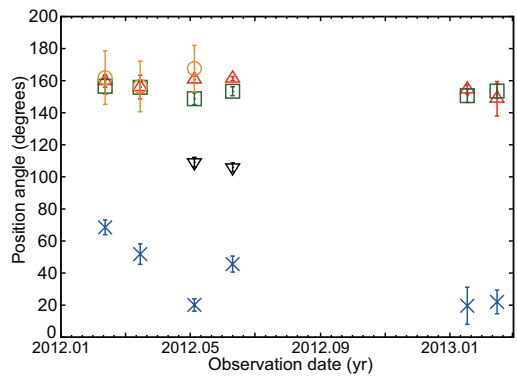}
\caption{Position angle of each component relative to the core. The same symbols as in Fig.~\ref{vlbarad} are used.}
\label{vlbapa}
\end{figure}

Figures~\ref{vlbarad} and \ref{vlbapa}
show the angular separation and the position angle of each component
relative to the core.
The components C1 and C2 are consistent with being quasi-stationary,
since there is no significant proper motion
by means of linear fits to the separation
from the core over more than three epochs
($0.02\pm0.15c$ for the component C1 and $-0.10\pm0.09c$ for the component C2).
We compare the locations of the components C1, C2, and C3 
(hereafter the southeast jet)
to those in previous studies.
\cite{Piner:2009} find that
VLBA images at 43\,GHz in 2005 have shown a clear limb-brightened structure to the southeast, 
which is labeled as the eastern limb ($PA\sim150^{\circ}$)
and western limb ($PA\sim175^{\circ}$)
at $0.5<r<2$\,mas
using the model fitting of elliptical Gaussians.
\cite{Piner:2010} find the eastern limb at a $PA$ of $\sim120^{\circ}$
and the western limb at a $PA$ of $\sim170^{\circ}$ during 2008--2009.
They also perform the model fitting of circular Gaussians as we do,
and identify components C4 and C5, 
which are located 0.58--0.72\,mas from the core at a $PA$ of 137$^{\circ}$--154$^{\circ}$
and 0.08--0.14\,mas at a $PA$ of $-170^{\circ}$--$-175^{\circ}$, respectively.
The component C2 in our images
could correspond to the component C4 in \cite{Piner:2010},
showing no detection of radial motion ($-0.01\pm0.01c$).
The averaged $PA$ of the components C1, C2, and C3 in our images is $\sim156^{\circ}$,
which is also comparable to the $PA$ of the component C4.
We find 
no emission at the location of the persistent western limb.

On the other hand, 
the characteristics of the NE component are different from
those of the southeast jet components.
It is difficult to perform model fitting using the relatively smooth emission distribution
around the NE component.
While the $PA$ of the NE component changes by $\sim50^{\circ}$ 
(from $\sim70^{\circ}$ to $\sim20^{\circ}$),
the $PA$s of the southeast components are contained within a narrower range 
($\sim15^{\circ}$).
The separations 
between the NE component and the core for the third and fourth epochs
are more than 0.1\,mas farther than those for the other epochs, 
whose separations are 
slightly smaller than the minor axis of the original beam size.
For the two epochs,
there might be difficulties in separating the core and the NE component
due to the apparent larger core sizes than the others. 
The absence of the long baselines (SC or MK)
could be one of the reasons for the apparent larger core sizes,
although the first epoch image can resolve the NE component from the core despite of the absence of SC baselines.
Therefore,
the apparent larger core sizes could be indeed intrinsic
due to a certain change of the physical conditions in the jet,
i.e., the Lorentz factor, the viewing angle, or the particle density in the jet.
\cite{Piner:2010} find the component C5 at a similar distance from the core ($\sim$0.11\,mas) in 2008 and 2009; 
however, its $PA$ ($\sim-170^{\circ}$)
differs by more than $100^{\circ}$ from that of the NE component.

\begin{figure}
\centering
\includegraphics[width=6cm]{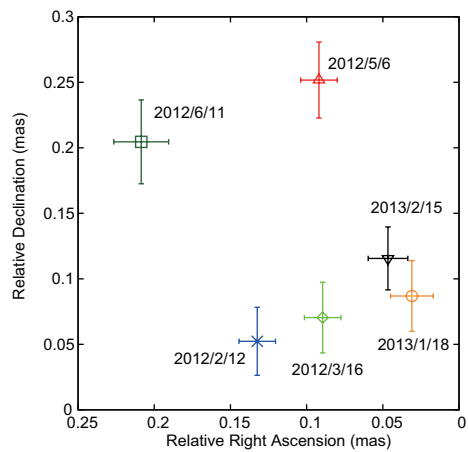}
\caption{Positions of the NE component relative to the core. The position of the core is set to (0,\,0). }
\label{twod}
\end{figure}

Figure~\ref{twod} shows that
the NE component does not follow any systematic motion.
The scattering of the positions of the NE component is within $\sim0.2$\,mas.
Except for the third and fourth epochs,
it could show a position change of $\sim0.1$\,mas to the northwest (or slightly inward).
Therefore, the characteristics of the NE component are different 
from those of regular $PA$ swings of the innermost jets 
(e.g., in the quasar NRAO~150, \citealt{Agudo:2007}; in the blazar OJ~287, \citealt{Agudo:2012}).

We also note that Mrk~501 shows several $PA$ changes 
out to the kiloparsec scale radio emission \citep{Giroletti:2008}.
The downstream jet $PA$ changes are always in the clockwise direction 
($\Delta PA=PA_{\rm out}-PA_{\rm in}<0$),
while in this case, the rotation is in the opposite direction, 
i.e., $\Delta PA=PA_{\rm C1}-PA_\mathrm{NE}>0$.
Thus, the $PA$ of the NE component does not follow the ``helical'' trajectory
of the outer jet,
as opposed to the curved jet in the blazar PKS~2136+141 \citep{Savolainen:2006a}.
Similar jet $PA$ mismatch between pc and kilo-pc scale is also observed in OJ~287 \citep{Agudo:2012}.
They suggest that different instability modes operate on the two spatial scales
and at smaller scales at higher frequencies ($\ge43$\,GHz)
the instabilities grow rapidly in amplitude.

\subsubsection{Flux density variability}~\label{FD}

\begin{figure}
\centering
\includegraphics[width=7cm]{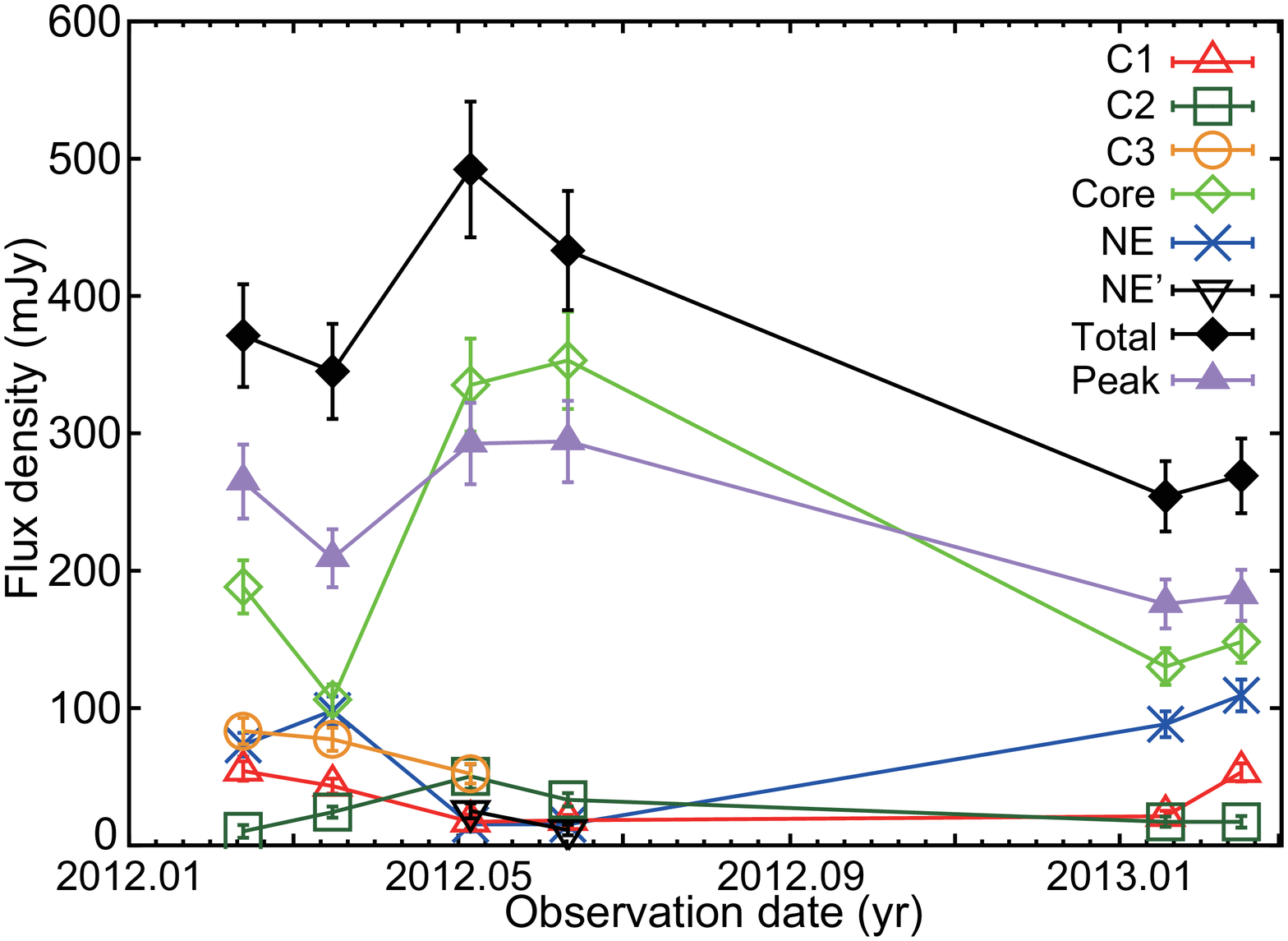}
\caption{Light curves of each component (C1, C2, C3, Core, NE, NE$'$), total CLEANed flux, and peak flux.}
\label{vlbalight}
\end{figure}

From the results of our model-fitting, 
we study the evolution of the flux density for each component at 43\,GHz. %
In Fig.~\ref{vlbalight},
we plotted the flux of all the components.
We perform a chi-square test of a flux variability for each component,
and define a flare to be an increase in the VLBI flux 
that is $>3\sigma$ of the averaged flux and greater than the noise level of the total flux density curve
\citep[e.g.,][]{Savolainen:2002}.
As a result, 
there are no significant variations in the flux densities of the components C2 and C3.
On the other hand, 
the flux densities of the component C1 are variable.
For the 1st, 2nd, and 6th epochs, the flux densities of the component C1 are 
twice as high as those for the other epochs.
These are more likely to be model-fit artifacts
due to the larger Gaussian components on faint and extended emissions.

Chi-square tests using the model-fit results on the flux densities in the core and the NE component 
also suggest that they are variable.
Although there are possible flux leakages between the closely spaced two components, 
when the core flux for the third and fourth epochs increased 150--250\,mJy,
the fluxes of the NE component decreased 60--90\,mJy.
Therefore,
an increase in the core flux is still significant,
such that there is a core flare during these two epochs.
This core flare could result from the larger core size, as is mentioned in \S\ref{AP}.
The total flux densities of two adjacent epochs (1\&2, 3\&4, 5\&6) are consistent within the error bars. 
The total fluxes of the third and fourth epoch increased by $\ge$100\,mJy,
and the increase in the total fluxes 
is consistent with the increase in the core flux.

In order to confirm the core flare,
we plotted the peak flux measured on images convolved 
with a circular beam of radius 0.3\,mas,
which is a conservative representation of the beam for the VLBA data at 43\,GHz \citep{Blasi:2013}.
For all epochs, 
the peak flux of the core with this circular beam
contains 60-70\% of the total flux density.
For the third and fourth epochs,
the peak flux is consistent with the core flux within the error bars.
For the other epochs, 
most of the flux densities of the core and the NE component are collected within a radius of about 0.3\,mas.
Nevertheless,
the peak fluxes are still variable,
such that we confirm the core flare.

\subsubsection{Slice profiles}\label{slpr}

\begin{figure}
\centering
\includegraphics[width=7cm]{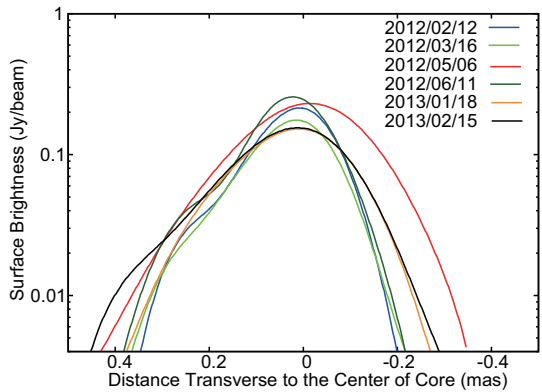}
\caption{Slice profiles of the brightness along the Gaussian center of the core and the NE component.
Location of the Gaussian center is set to the reference position.
All the profiles bulge eastward.}
\label{vlbaslice}
\end{figure}

To emphasize the existence of the NE component,
we made a slice profile of the CLEAN image
across the center of Gaussian models of the core and the NE component 
at each epoch
using the \AIPS 
~task {\sc slice}.
In Fig.~\ref{vlbaslice} we show
all the slice profiles 
referenced to the Gaussian center of the core.
Every slice profile is asymmetric and  shows the presence of two humps
corresponding to the core and the NE component.
We fitted both single and double Gaussian component models to each slice profile
using the \AIPS 
~task {\sc slfit}.
Table~\ref{sltab} lists the parameters of the fitted Gaussian models.
By comparing the rms of the residuals for the single and double Gaussian fits,
we determine that double Gaussian components provide a better fit.
Although the parameters of the Gaussian models for the NE component  
are different from those in Table~\ref{gaus}
owing to the difference of the dimensions,
we confirm that there is clear emission at the northeast of the core for all epochs
with a peak flux of $\sim36\,{\rm mJy~beam^{-1}}$ 
at the position of $\sim0.18$\,mas from the core
and a FWHM of $\sim0.24$\,mas on average.

\begin{table}
\caption{Gaussian models for slice profiles of VLBA 43\,GHz images}
\label{sltab}
\resizebox{0.80\textwidth}{!}{\begin{minipage}{\textwidth}
\begin{tabular}{ccccccccccc}
\hline\hline
Epoch & Model\tablefootmark{a} & ID & Peak & Position & FWHM & rms\tablefootmark{b}\\
&&& (${\rm mJy~beam^{-1}}$) & (mas) & (mas) & (mJy) \\
\hline
BK172A &        1 &     Core & 208 & ... & 0.20 & 158\\
        &       2 &     Core     & 207 & ... & 0.17 & 25\\
        &       2 &     NE       & 39  & 0.16 & 0.21\\  
BK172B &        1 &     Core & 169 & ... & 0.22 &       123\\
        &       2 &     Core & 164 & ... & 0.18 & 22\\
        &       2 &     NE       & 34 & 0.15 &  0.24\\  
BK172C &        1 &     Core & 227 & ... & 0.31 &       93\\
        &       2 &     Core & 215 & ... & 0.27 &       29\\
        &       2 &     NE & 36  & 0.18 & 0.30\\        
BK172D &        1 &     Core & 247 & ... & 0.20 &       194\\
        &       2 &     Core & 254 & ... & 0.19 &       28\\
        &       2 &     NE      & 38 & 0.22 &   0.15\\  
BK172E &        1 &     Core & 150 & ... & 0.28 &       70\\
        &       2 &     Core & 137 & ... & 0.23 &       16\\
        &       2 &     NE & 40 & 0.16 & 0.24\\ 
BK172F &        1 &     Core & 151 & ... & 0.29 &       107\\
        &       2 &     Core & 145 & ... & 0.25 &       19\\
        &       2 &     NE & 29 & 0.20 & 0.30\\ 
\hline
\end{tabular}
\\
\tablefoottext{a}{Fitting model type. Model 1 represents a single Gaussian model fitting. \\
Model 2 represents a double Gaussian model fitting.} \\
\tablefoottext{b}{Root sum squares of the residuals.}                                                                                                                                                                                                                                                                                                                                                                                                                                                                                                                                                                                                                                                                                                                                                                                                                                                                                                                                                                                                                                                                                                                                                                                                                                                                                                                                                                                                                                                                                                                                                                                                                                                                                                                                                                                                                                                                                                                                                                                                                                                                                                                                                                                                                                                                                                                                                                                                                                                                                                                                                                                                                                                                                                                                                                                                                                                                                                                                                                                                                                                                                                                                                                                                                                                                                                                                                                                                                                                                                                                                                                                                                                                                                                                                                                                                                                                                                                                                                                                                                                                                                                                                                                                                                                                                                                                                                                                                                                                                                                                                                                                                                                                                                                                                                                                                                                                                                                                                                                                                                                                                                                                                                                                                                                                                                                                                                                                                                                                                                                                                                                                                                                                                                                                                                                                                                                                                                                                                                                                                                                                                                                                                                                                                                                                                                                                                                                                                                                                                                                                                                                                                                                                                                                                                                                                                                                                                                                                                                                                                                                                                                                                                                                                                                                                                                                                                                                                                                                                                                                                                                                                                                                                                                                                                                                                                                                                                                                                                                                                                                                                                                                                                                                                                                                                                                                                                                                                                                                                                                                                                                                                                                                                                                                                                                                                                                                                                                                                                                                                                                                                                                                                                                                                                                                                                                                                                                                                                                                                                                                                                                                                                                                                                                                                                                                                                                                                                                                                                                                                                                                                                                                                                                                                                                                                                                                                                                                                                                                                                                                                                                                                                                                                                                                                                                                                                                                                                                                                                                                                                                                                                                                                                                                                                                                                                                                                                                                                                                                                                                                                                                                                                                                                                                                                                                                                                                                                                                                                                                                                                                                                                                                                                                                                                                                                                                                                                                                                                                                                                                                                                                                                                                                                                                                                                                                                                                                                                                                                                                                                                                                                                                                                                                                                                                                                                                                                                                                                                                                                                                                              
\end{minipage}}
\end{table}

\subsection{The GMVA results}\label{gmvares}

\begin{figure}
\centering
\includegraphics[width=6cm]{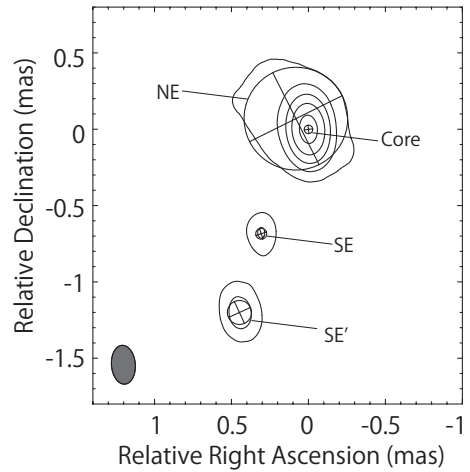}
\caption{The GMVA image of Mrk~501 at 86\,GHz with four circular Gaussian model components 
with the restoring beam of 255~$\mu$as $\times$ 155~$\mu$as in $PA~4.41^{\circ}$.
The peak brightness is 235\,${\rm mJy~beam^{-1}}$, and the contour levels are drawn at (-1,~1,~2,~4,…)$\,\times\,10.7\,{\rm mJy~beam^{-1}}$. The 1$\sigma$ noise level is $3.6\,{\rm mJy~beam^{-1}}$.}
\label{gg072}
\end{figure}

In Fig.~\ref{gg072},
we show the scaled GMVA image of Mrk~501 
with Gaussian models.
Visibility model-fitting in Difmap provides a reduced $\chi^{2}=0.86$ with this model.
The result of model-fitting is summarized in Table~\ref{gaus}.
Owing to the lack of sensitivity on long baselines and sparse ($u,v$) coverage,
the visibility data are weighted using the natural weighting and tapered
using a Gaussian taper with $\sigma_{\rm taper}=1000~{\rm M\lambda}$.
This yields a synthesized beam of $255~{\rm \mu as} \times 155~{\rm \mu as}$,
which is about ten times larger than the beam obtained by \cite{Giroletti:2008},
and comparable to the beam size in Fig.~\ref{res01835}.
We find an extended core emission modeled by a circular Gaussian,
which is located 0.11\,mas northeast of the core with $\sim$0.7\,mas in radius.
Since the location of the Gaussian model is comparable to those of the NE component in 43\,GHz images,
we label this component NE.
We also find a jet knot (labeled SE) located at $(r,~\theta)=(0.75~{\rm mas}, 156^{\circ})$.
This knot could correspond to the one detected by the GMVA at 86\,GHz in \cite{Giroletti:2008},
located at $(r,~\theta)=(0.73~{\rm mas}, 172^{\circ})$.
The angular separation of the southeast knot from the core is comparable between these two epochs,
which is consistent with the absence of superluminal jet motion.
However, since there is a clear $PA$ difference ($\sim16^{\circ}$),
the SE component  could be different from the knot in \cite{Giroletti:2008}.
An extended jet knot (SE$'$) located at $(r,\theta)=(1.28~{\rm mas}, 160^{\circ})$
could correspond to  component C1 in the 43\,GHz images.

Owing to low S/N of the visibilities at transatlantic baselines,
we set the upper limit of the unresolved core size as $55~\mu$as
(0.037~pc) in radius,
which is the deconvolved angular size of the core.
Since we could not perform the amplitude self-calibration because of low S/N, we estimate the flux density uncertainties considering 
both a conservative flux calibration error of about 30\%
at 86\,GHz \citep{Lee:2008} and three times the map rms noise.

\subsection{VLBA 43\,GHz vs. GMVA 86\,GHz}
\subsubsection{Brightness temperature}\label{bt}

We derive a lower limit of the brightness temperature in the core at 86\,GHz
using an upper limit of the angular size and a lower limit of the flux.
The brightness temperature for the core is estimated
to be $T_{B}\ge4.5\times10^{9}$~K.
Although the flux of the core is more than three times as high as
those of the previous 86\,GHz observations \citep{Giroletti:2008,Lee:2008},
the brightness temperature is comparable to their values
owing to the larger deconvolved size.
Therefore, it does not require a high Doppler factor at the base of the radio jet.
For the 43\,GHz images,
we derive the brightness temperature in the core
using the deconvolved core size.
The brightness temperatures in the core at 43\,GHz
are estimated
to be around a few $\times10^{9}$~K,
which is consistent with the values in \cite{Piner:2010} at 43\,GHz,
and comparable to the $T_{\rm B}$ at 86\,GHz.

The brightness temperatures of the NE component
are
$T_{B}\sim10^{8}$~K for VLBA 43\,GHz
and
$T_{B}\ge3\times10^{7}$~K for GMVA 86\,GHz,
which are much lower than that of the core
and do not require a high Doppler factor.
%

\subsubsection{Spectral index}\label{si}

\begin{figure*}
\centering
\includegraphics[width=13cm]{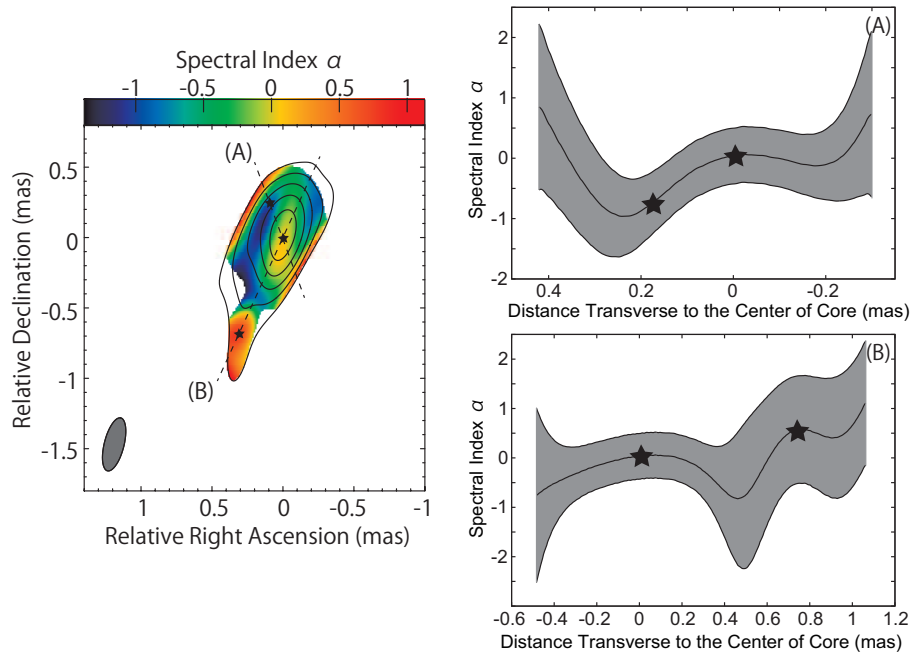}
\caption{{\it Left panel}: Spectral index map between the third epoch VLBA 43\,GHz map and the GMVA 86\,GHz map.
The contour map shows the VLBA 43\,GHz map,
with the contour levels drawn at (-1,~1,~2,~4,…)$\,\times\,9.0\,{\rm mJy~beam^{-1}}$,
which is three times the rms noise of the off-source region of the 86\,GHz image.
Black dotted lines show the locations of the slices
and black stars indicate the centroid positions of the fitted Gaussian models. 
{\it Right panel}: (A) Slice profile of spectral index map shown in Fig.~\ref{spix} along Gaussian center of the core and the NE component  at $PA=20.1^{\circ}$ for the third epoch 43\,GHz observations.
The gray area indicates $1\sigma$ errors on the spectral index.
The black stars show locations of the Gaussian models for the core and the NE component 
in the third epoch 43\,GHz map.
(B) Slice profile of spectral index map shown in Fig.~\ref{spix} along Gaussian center of the core and the SE component  at $PA=155.8^{\circ}$ for the GMVA 86\,GHz observations.
The black stars show locations of the Gaussian models for the core and the SE component  
in the 86\,GHz map.}
\label{spix}
\end{figure*}

In the left panel of Fig.~\ref{spix},
we show the spectral index distribution between
the third epoch VLBA 43\,GHz image and the GMVA 86\,GHz image,
which were obtained over a period of two weeks.
We define the spectral index $\alpha$ as $S_{\nu}\propto\nu^{\alpha}$.
The same ($u,v$) range (32--668~M$\lambda$) and the same restoring beam size are used
for the images of both frequencies.
Then we produced the spectral index map
by using the \AIPS 
~task {\sc comb},
clipping the pixels with three times the rms noise of the off-source region of the 86\,GHz image,
referencing to the Gaussian center of the core for both images.
Given the short time separation between these two images,
we can obtain quasi-simultaneous spectral index maps.
From this figure,
we find the core region has a flat spectrum,
the NE component has a flat-to-steep spectrum,
and the southeast jet region has a flat-to-inverted spectrum.
We made spectral index maps 
by taking into account the core shift effect
($\sim30~\mu$as between 43\,GHz and 86\,GHz, 
extrapolated from the power-law relation between the core position and frequency of Mrk~501 reported in \citealt{Croke:2010}).
However, the spectrum did not change significantly.
Although we also made spectral index maps
by using the 86\,GHz map without flux scaling,
the spectral indices are consistent with the values with flux scaling within the error bars
and the spectral tendency did not change.

Since the positions of Gaussian models between 43\,GHz and 86\,GHz
do not coincide with each other,
we estimate the value of the spectral indices 
using the slice profiles of the spectral index distribution \citep[e.g.,][]{Hovatta:2014}.
The uncertainties of the spectral index 
were calculated from the theory of the propagation of errors,
including the frequency-dependent flux-density uncertainty
and three times the map rms noise \citep[e.g.,][]{Lico:2012}.
The edges of the spectral index profiles
have larger spectral uncertainties
due to lower flux density than at the center of the profiles.
The top right panel of Fig.~\ref{spix} shows
the spectral index profile along the Gaussian center of the core
and the NE component position at the third epoch 43\,GHz observations ($PA=20.1^{\circ}$).
The bottom right panel shows 
the spectral index profile along the Gaussian center of the core
and the SE component position at the 86\,GHz observations ($PA=155.8^{\circ}$).
From the spectral index profiles,
we find that
the spectral index of the core is $\alpha_\mathrm{c}=0.0\pm0.5$,
that of the NE component  is $\alpha_\mathrm{NE}=-0.8\pm0.5$,
and that of the SE component  is $\alpha_\mathrm{SE}=0.6\pm1.1$.
The spectral index of the core is consistent with
previous results
($\alpha\sim-0.5$ above the turnover, \citealt{Giroletti:2008};
$\alpha\sim0.2$ between 8\,GHz and 15\,GHz, \citealt{Hovatta:2014}).
Compared with the core,
the NE component has a steeper spectral index, and
the SE component has a comparable or flatter spectral index.
The flatter spectrum at the edges of the spectral profile in Fig.~\ref{spix} (A)
is likely an artifact due to the large Gaussian size of the NE component the 86\,GHz image.

%

\section{Discussion}
\subsection{Possible origins of the NE component}\label{VL}

Since the NE component does not follow any systematic motion (\S\ref{AP}),
it can be regarded as a quasi-stationary component.
Here we discuss two possible origins of the quasi-stationary component,
which are internal shocks among discrete blobs 
and a part of dim continuous flow.

\subsubsection{Internal shocks}

\begin{figure}[htbp]
\begin{center}
\includegraphics[width=6cm]{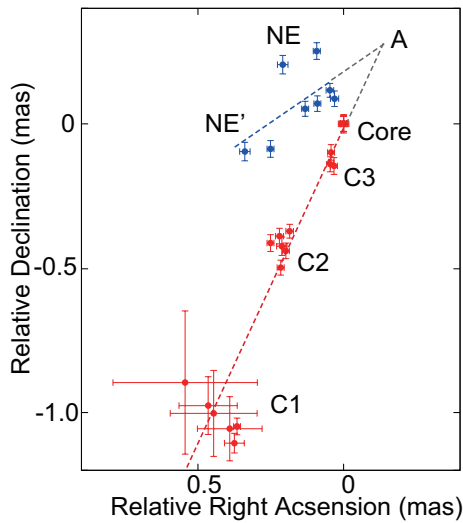}
\caption{Blue dots show the positions of the components NE and NE$'$, and red dots show the positions of the radio core, the components C1, C2, and C3 for all epochs.
We define the dotted blue line (the least square line of the blue dots)
as the eastern limb,
and the dotted red line (that of the red dots)
as the western limb.
The intersection of the extrapolated lines (shown as the dotted grey lines) is labeled A.
The position of the core is set to (0,\,0).}
\label{NEfit}
\end{center}
\end{figure}

One of the possible origins of the NE component
is internal shocks.
The internal shock model is one of the standard models for explaining blazar emission 
\citep[e.g.,][]{Spada:2001fj,Tanihata:2003,Koyama:2015}.
Based on the internal shock model,
variation of Lorentz factors of the ballistic discrete ejecta makes
the distance between the internal shocks and the central engine different.
When we also take into account the effect of changing jet axes alignment as well,
both the NE component and the radio core can be explained
as the internal shock regions
along with different jet axes.
Two jet axes are determined by least square fitting
to the component positions for all the epochs in Fig.~\ref{NEfit};
one axis is fitted to the component NE and NE$'$ (eastern limb as the dotted blue line),
and the other axis is fitted to the core, 
and the components C1, C2, and C3 (western limb as the dotted red line).
In this case, the origin of the flow,
which corresponds to the intersection of the two axes, 
is expected to be in the upstream of the core
(Fig.~\ref{NEfit} A).
The angular separation between the intersection of the two axes and the NE component is
$0.3$\,mas, or 2.8\,pc de-projected 
by assuming the jet viewing angle of $\sim4^{\circ}$ \citep[e.g.,][]{Giroletti:2008}.
The distance between the central engine
and the location of the internal shock region
is approximately proportional to the square of the Lorentz factor of the ejecta \citep[e.g.,][]{Spada:2001fj}.
Hence, 
the quasi-stationary location of the internal shock regions
can be explained by a small variation of Lorentz factors.
In fact, the expected distance from the central engine to the NE component
is comparable to the distance to the core in \cite{Koyama:2015}
where we explained that a small variation of Lorentz factors in its quiescent state
makes the spatial distribution of the location of the internal shocks within 0.2\,mas.
Therefore, internal shocks can be one of the possible origins of the quasi-stationary NE component.

It is also worth discussing what will happen 
in its active state.
When the variation of the Lorentz factor is significant in its active state,
the location of the internal shock region
would change considerably.
In the case of another nearby TeV blazar Mrk~421,
we detected significant core wandering up to 0.5\,mas
soon after its big X-ray flare \citep{Niinuma:2015}.
If the radio core wanders $\sim0.5$\,mas 
when Mrk~501 is in active state,
the location of the NE component would spread widely, or
it might show systematic motion.
We need further constraints on the core position by using astrometric observations
to study the exact location of the NE component.

\subsubsection{A part of dim flow}

Another possible origin of the NE component 
is a part of dim continuous flow,
which is usually below the detection threshold of VLBA at 43GHz.
Under certain conditions,
such as the injection of radio-emitting plasma at the base of the jet 
related to a high-energy flare, 
or a slight increase in Doppler factors,
the observed flux can be enhanced.
\footnote[4]{If the enhanced flux of the NE component
is related to a high-energy flare,
it could be a $\gamma$-ray flare in 2011 October \citep{Bartoli:2012} (\S\ref{IM}).}
Then the enhanced emission can be above the detection threshold.
The possible flux variability of the NE component (\S\ref{FD})
may support this scenario.
Although there are possible flux leakages between the core and the NE component,
the moderate variability (from 15\,mJy to 109\,mJy over $\sim8$ months with the size of 0.26\,mas) suggests
a low value of ``variability Doppler factor''
\footnote[5]{defined by \cite{Jorstad:2005} 
by assuming that the variability timescale corresponds to the light-travel time across the knot}
as $\delta\sim3$.
This value is consistent with the results 
obtained by 
low brightness temperatures obtained in \S\ref{bt}.
The location of the NE component is quasi-stationary within $\sim0.2$\,mas over one year,
which puts the upper limit of its apparent velocity as $\beta_{\rm app}\la0.4$
(0.10\,mas~yr$^{-1}=0.22c$).
If we adopt the jet viewing angle of $\theta\sim4^{\circ}$ \citep[e.g.,][]{Giroletti:2004},
the corresponding Lorentz factor of the emission region is estimated to be $\Gamma\la2$
(a Doppler factor of  $\delta\la4$).
Since the obtained upper limit of the Lorentz factor is 
lower than the typical values of one-zone SED modeling
(e.g., $\delta\sim\Gamma\sim10$: \citealt{Bartoli:2012}),
the NE component is not completely identical to 
the high-energy emission region.
Such a low value of the Lorentz factor may support the idea
that the NE component corresponds to the slower outer layer.

Finally, we discuss the origin of the dim flow.
The origin of the dim flow can be the core or the intersection of the two axes
(position A shown in Fig.~\ref{NEfit}).
If the origin of the jet is located at the core,
the jet should be initially ejected toward the northeast direction
from the core rather than the persistent southeast jet,
then it apparently bends $\sim90^{\circ}$ toward the southeast.
New component emergence at an unusual $PA$
similar to Mrk~501 has been reported in the quasar 3C~279;
however,
it does not show quasi-stationary behavior like the NE component in Mrk~501 
but shows an outward motion \citep{Lu:2013}.
Such an unusual jet $PA$ could be similar to
the erratic $PA$ changes of innermost jets, which is possibly caused 
by hydrodynamic instabilities due to asymmetric jet injection
\citep[e.g.,][]{McKinney:2009aa,Aloy:2003},
as observed in a few blazars
(e.g., OJ~287: \citealt{Agudo:2012}, BL Lac: \citealt{Cohen:2014}).
However,
it is difficult to bend the jet apparently $\sim90^{\circ}$,
that is, intrinsically $\sim40^{\circ}$ (for $\theta\sim4^{\circ}$)
\citep[e.g., via oblique shocks:][]{Tingay:1997}.
Therefore,
the upstream of the core (position A in Fig.~\ref{NEfit})
is more likely to be the jet origin,
and a slight change in the jet direction toward observers at the NE component
can enhance the Doppler factor and the flux of the dim flow
\citep[e.g.,][]{Gomez:1993}.

\subsection{Spectral index of southeast jet feature}\label{SID}

Our observation could not provide strong constraints on the spectral index of the jet feature because of the limited 
capability of the array.
To obtain more precise GMVA images,
we need better ($u,v$)-coverage
to be provided by the addition of new stations to the array.
Recently, the Korean VLBI Network \citep{Lee:2014a}
joined GMVA observations \citep{Hodgson:2014}, which will increase the number of east-west baselines
when there is no common sky between Europe and the USA.
The east-west baselines of $\sim9000$~km
is crucial for resolving the core and the NE component 
with typically 50~$\mu$as resolution.
Filling short baseline spacings will also
enable us to perform more accurate calibration.
Thus, the increased number of the baselines
is important for obtaining images and spectral index maps at higher fidelity.

Between 8 and 15\,GHz,
the spectral index at the component 0.6\,mas southeast of the core of Mrk~501
is $\alpha=-0.5\pm0.1$,
and the mean spectral index of the overall jets of BL Lac objects is $\alpha\sim-0.8$
\citep{Hovatta:2014}.
They also point out that the jet spectrum flattens on average by $\sim0.2$
at the location of the jet components,
indicating the enhancement of particle density
due to shocks in the jets \citep{Mimica:2009},
or superposition of multiple components along the line of sight.
A flat-to-inverted spectral distribution is also seen at the component C4,
located 0.6\,mas southeast of the core, 
having $\alpha\sim0.2$ between 1.6\,GHz and 4.8\,GHz space VLBI observations \citep{Giroletti:2004},
while the other components have a steep spectrum.
\cite{Edwards:2000} also suggest that it could be connected to the inner spine flat-spectrum region.
If the SE component is a part of the inner spine,
the location of the component could change $\sim0.3$\,mas 
within the separation period between 43\,GHz and 86\,GHz observations (13 days),
by applying $\theta_\mathrm{c}=4^{\circ}$ and $\Gamma_\mathrm{c}=17$, so that it leads to large image misalignment.
Among other sources,
a flat-to-inverted spectral distribution is seen in one of the index maps of
the jet in the quasar NRAO~150 between 43\,GHz and 86\,GHz \citep{Molina:2014},
but the separation of four months between the two images
may cause large image misalignment due to structure change.
Since such a flat-to-inverted jet spectrum is quite unusual
and has not yet been examined precisely,
it is important to confirm whether it is real 
by performing simultaneous observations between 43 and 86\,GHz. 


\section{Conclusions}

We have explored the sub-pc-to-pc scale region in Mrk~501 
using the VLBA at 43\,GHz and the GMVA at 86\,GHz
between 2012-2013.
   \begin{enumerate}
      \item We find a new off-axis jet structure, the NE component, located northeast of the core.
      The position angle of the NE component changes from $70^{\circ}$ to $20^{\circ}$,
      while those for the southeast jet components reside within $\sim15^{\circ}$.
      The scattering of the positions of the NE component relative to the core
      is within $\sim0.2$\,mas
      for all of our VLBA 43\,GHz observations.
      There is a small core flare during the third and fourth epochs,
      and it could be due to the large core size.
      From the slice profiles, 
      the component NE is located at $0.18$\,mas with the peak flux of $36\,{\rm mJy~beam^{-1}}$
      and FWHM of $0.24$\,mas on average.
      \item We obtain the quasi-simultaneous GMVA 86\,GHz image 
      within two weeks of the third epoch of the VLBA 43\,GHz observations. 
      Although the visibilities on transatlantic baselines show weak coherence and low signal-to-noise,
      we detect the northeast feature located at 0.11\,mas from the core, and
      the southeast jet feature located at 0.75\,mas and 1.28\,mas from the core.
      \item From the spectral index profile analysis between the quasi-simultaneous 43 and 86\,GHz images,
      the spectral index of the core is flat and consistent with previous results,
      while that of the NE component is flat-to-steep, and that of the SE component is flat-to-inverted.
      To confirm these values, 
      simultaneous high-frequency observations with more east-west baselines are necessary.
      \item 
      Since the NE component can be regarded a quasi-stationary component,
      internal shocks or a part of dim flow
      could be possible origins of the component.
      Determination of the more precise location of the NE component using radio core astrometry,
      and higher sensitivity observations toward the dim flow are needed.    
      \end{enumerate}

\begin{acknowledgements}
We thank the anonymous referee for useful comments and suggestions.
The National Radio Astronomy Observatory is a facility of the National Science Foundation operated under cooperative agreement by Associated Universities, Inc. This work made use of the Swinburne University of Technology software correlator \citep{Deller:2011}, developed as part of the Australian Major National Research Facilities Programme and operated under license.
This paper is partially based on observations carried out with the VLBA, the MPIfR 100\,m Effelsberg Radio Telescope, the IRAM Plateau de Bure Millimetre Interferometer, the IRAM 30\,m Millimeter Telescope, the Onsala 20\,m Radio Telescope, and the Mets\"{a}hovi 14\,m Radio Telescope.
IRAM is supported by MPG (Germany), INSU/CNRS (France), and IGN (Spain). The GMVA is operated by the MPIfR, IRAM, NRAO, OSO, and MRO. 
We thank the staff of the participating observatories for their efficient and continuous support.
S.K. acknowledges this research grant provided 
by the Global COE program of University of Tokyo.
Part of this work was done with the contribution of the Italian Ministry of Foreign Affairs and University and Research for the collaboration project between Italy and Japan.
This work was partially supported by Grant-in-Aid
for Scientific Research, KAKENHI 24340042 (A.D.) and 2450240 (M.K.) from the
Japan Society for the Promotion of Science (JSPS).
E.R. and M.A.P.T. acknowledge partial support from the Spanish MINECO through projects AYA-2012-38491-C02-01 and AYA-2012-38491-C02-02. E.R. also acknowledges partial support from the Generalitat Valenciana through project PROMETEOII/2014/057.
S.K. is thankful to Jeffrey A. Hodgson and Bindu Rani for useful discussion and comments on the paper.
      
\end{acknowledgements}

\bibliographystyle{aa}
\bibliography{VLBAGMVA501.bib}

\end{document}